\documentclass[unicode, 12pt, a4paper, twoside]{article}
\usepackage[english]{babel}

\usepackage{vmargin}

\usepackage[style=numeric-comp, sorting=none, url=false, doi=false, firstinits=true, backend=bibtex, eprint=false, isbn=false]{biblatex}
\bibliography{Literature.bib}

	\usepackage{graphicx}
	\usepackage{color}
		\definecolor{emphblue}{rgb}{0.3,0.3,0.7}
		\renewcommand{\emph}[1]{\textcolor{emphblue}{\textbf{#1}}}
		
	\setmarginsrb{3 cm}{2 cm}{2 cm}{2 cm}{0 pt}{0mm}{0pt}{13mm}

	\usepackage{url}
	\usepackage[unicode=true]{hyperref}

	\usepackage{authblk}  
	\usepackage{units}
	\usepackage{cool} 

\title{Data transmission in long-range dielectric-loaded surface plasmon polariton waveguides}

\author[a,*]{\small \bf Svyatoslav Kharitonov\,}
\author[a,*]{\small \bf Roman Kiselev\,}
\author[a]{\small \bf Ivan Fern\'andez de J\'auregui Ruiz\,}
\author[b]{\small \bf Ashwani Kumar\,}
\author[b,c]{\small \bf Xueliang Shi\,}
\author[d]{\small \bf Kristj\'an Le\'osson\,}
\author[a]{\small \bf Thomas Pertsch\,}
\author[b]{\small \bf Sergey Bozhevolnyi\,}
\author[a]{\small \bf Stefan Nolte\,}
\author[a]{\small \bf Arkadi Chipouline\,}

\affil[a]{\scriptsize Institute of Applied Physics, Friedrich-Schiller University Jena, Albert-Einstein Str. 15, D-07745 Jena, Germany}
\affil[b]{\scriptsize Department of Technology and Innovation, University of Southern Denmark, Niels Bohrs All\'e 1, DK-5230 Odense M, Denmark}
\affil[c]{\scriptsize Department of Information Science and Electronic Engineering, Zhejiang University, Hangzhou 310027, China}
\affil[d]{\scriptsize Department of Physics, Science Institute, University of Iceland, Dunhaga 3,
IS-107 Reykjavik, Iceland}
\affil[*]{\scriptsize These authors contributed equally to the work}

\begin{document}
\maketitle
\section{Introduction}

Optical components, exploiting properties of surface plasmon polaritons (SPP) as guided waves, unlock one of the ways to on-chip photonics, which demands high data bandwidth, low power consumption, and small footprint of the integrated circuits \cite{Sorger2012}. SPP represents a surface wave, propagating along a boundary between two media, which possess opposite signs of the real part of permittivity (for instance, metal at the frequencies lower than the plasmonic frequency and dielectric). Electrons on a metal boundary can perform coherent with electromagnetic field collective motion \cite{Ritchie1957}, and thus support a strong mode confinement at the subwavelength scale, which is an attractive feature allowing to scale down the size of integrated optical circuitry.

Various of nanoplasmonic devices starting from passive components like waveguides \cite{Kalavrouziotis2012, Ju2007},  couplers and polarization beam combiners \cite{Zenin2012, Melikyan2012}, to active components, e.g. ring and disk resonators \cite{Randhawa2011, Holmgaard2009, Krasavin2010}, and modulators \cite{Pacifici2007, Melikyan2011, Krasavin2011}
have been recently demonstrated. Authors of \cite{Kalavrouziotis2012} reported a successful transmission of 480 Gbit/s aggregated data traffic (12 channels $\times$ 40 Gbit/s) over dielectric loaded plasmonic waveguides.

Dielectric-loaded SPP waveguides (DLSPPWs) consist of a dielectric stripe deposited on top of a metallic film and represent one of the basic configurations of the plasmonic waveguides \cite{Holmgaard2007}. DLSPPWs demonstrate sub-wavelength confinement of SPPs with the typical propagation distance of $\sim 50\, \unit{\mu m}$ at $\sim 1550\,\unit{nm}$ wavelength \cite{Holmgaard2008}. The propagation length can be significantly improved in the so-called long-range DLSPPWs (LR-DLSPPWs) that provide both mm-long SPP guiding and relatively tight mode confinement \cite{Holmgaard2010, Gosciniak2011}. Experimental investigations of LR-DLSPPWs operating at continuous-wave telecommunication band signal ($\sim 1550\,\unit{nm}$) demonstrated propagation length $\sim 500\,\unit{\mu m}$ with the mode size $\sim 1\,\unit{\mu m}$ \cite{Volkov2011}. In this paper we report successful transmission of 10 Gbit/s on-off-keying (OOK) modulated signal through the LR-DLSPPWs with almost negligible degradation of the data flow consistency.

\section{Data transmission experiment}

\subsection{Description of the tested waveguides}

\begin{figure}[h!]
\centering
\includegraphics[width=1\linewidth]{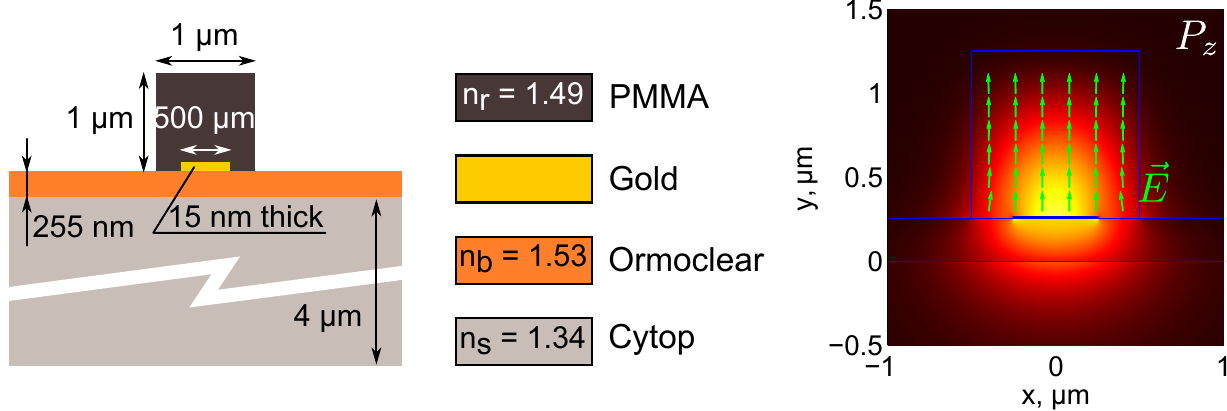}
\caption{Layout of the LR-DLSPP waveguide structure profile and distribution of the supported TM-mode at the wavelength 1550 nm (calculated with COMSOL Multiphysics v4.4)}
\label{fig:wg_profile}
\end{figure}

The detailed description of similar waveguides is presented in \cite{Volkov2011}. The waveguiding structure is fabricated by spin-coating of a 255\,nm thick layer of UV-curable organic-inorganic hybrid material (Ormoclear, $n_b$~= 1.53) onto a 4 $\mu$m thick amorphous fluoropolymer (Cytop, $n_s$~= 1.34) coated SiO$_2$ wafer. On the top of the Ormoclear layer there are $\sim 15\,\unit{nm}$ gold stripes covered with PMMA ($n_r$~= 1.49). The purpose of the gold stripe is to provide a strong confinement of the electromagnetic wave. The width of the produced PMMA ridges is close to 1 $\mu$m, which ensures the single-mode operation. The structure of the waveguide is depicted in Fig.~\ref{fig:wg_profile}.

The investigated waveguide has length of 300 $\mu$m. One side of the optical chip with patterned DLSPPWs was cleaved. The opposite side of the waveguide is terminated by a DLSPP taper with a diffraction grating etched on it. The grating scatters light at angle of 15$^\circ$ relative to the vertical direction.

\subsection{Prototype of the transmission line}

\begin{figure}[h!]
\centering
\includegraphics[width=1\linewidth]{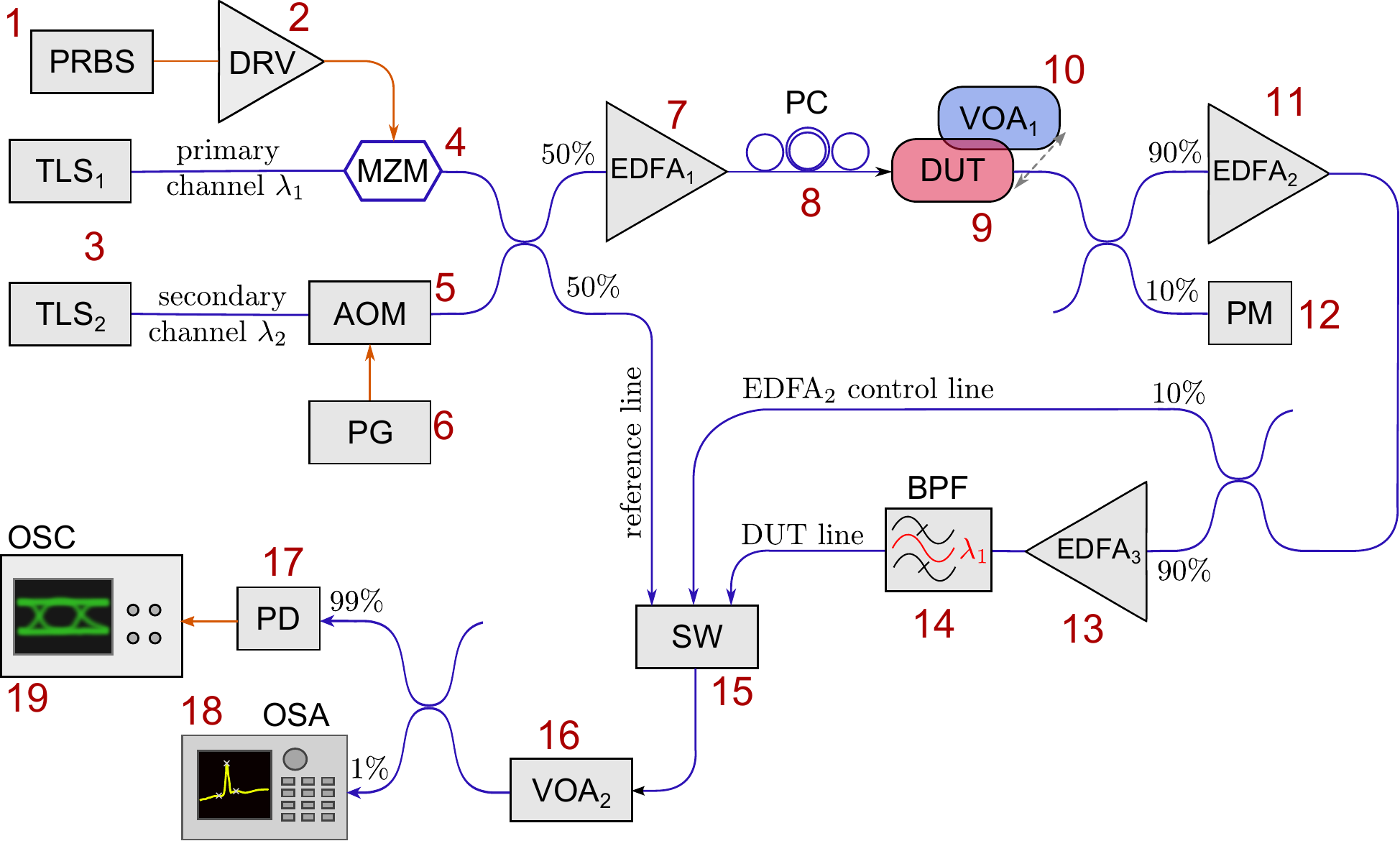}
\caption{Scheme of the experimental setup for the BER measurements:
	        \emph{1}~--- pseudorandom bit sequence generator (\emph{PRBS});
	        \emph{2}~--- electrical driver (\emph{DRV});
	        \emph{3}~--- tunable laser source (\emph{TLS$_{1,2}$});
	        \emph{4}~--- Mach-Zehnder modulator (\emph{MZM});
	        \emph{5}~--- Acousto-optical modulator (\emph{AOM});
	        \emph{6}~--- pattern generator (\emph{PG});
	        \emph{7, 11, 13}~--- erbium-doped fiber amplifiers (\emph{EDFA$_{1,2,3}$});
	        \emph{8}~--- polarization controller (\emph{PC});
	        \emph{9}~--- device under test (\emph{DUT});
	        \emph{10}, \emph{16}~--- motorized variable optical attenuators (\emph{VOA});
	        \emph{12}~--- power meter (\emph{VOA});
	        \emph{14}~--- tunable band-pass filter (\emph{BPF});
	        \emph{15}~--- fiber optic switch (\emph{SW});
	        \emph{17}~--- fast photodetector (\emph{PD});
	        \emph{18}~--- optical spectrum analyzer  (\emph{OSA});
	        \emph{19}~--- sampling oscilloscope (\emph{OSC}).
	         }
\label{fig:experimental_setup}
\end{figure}

In the performed tests, a typical prototype of a fiber optic telecommunication line, consisting of transmitter (blocks \emph{1}--\emph{8}), receiver (\emph{11}--\emph{19}) and device under test (\emph{DUT})(\emph{9}--\emph{10}), was used (see Fig.~\ref {fig:experimental_setup}).

The transmitter unit includes two tunable, independently modulated laser sources (TLS), multiplexed using a 50/50\% coupler and pre-amplified using an erbium-doped fiber amplifier (EDFA$_1$). A laser of the primary channel (CH1) is modulated using a Mach-Zehnder modulator (MZM). The MZM is driven with 10 Gbit/s electrical signal that represents a pseudorandom bit sequence (PRBS) with a length of $2^{32}-1$ bits. A secondary channel (CH2) laser is modulated with a low-rate 2 Mbit/s square wave signal using an acousto-optical modulator (AOM). The total output power of EDFA$_1$ is kept constantly at the level of 20 dBm to prevent a thermal damage of the waveguides.
A polarization controller (PC) is used to ensure the TM polarization of the light, injected into LR-DLSPPW.

After a propagation through the DUT, a weak signal is amplified by means of two cascaded EDFAs (\emph{11, 13}). An optical band-pass filter (BPF) (0.2 nm FWHM) is used to demultiplex the CH1 signal. Then, a variable optical attenuator (VOA$_2$) reduces the signal down to the level not exceeding a saturation limit of the photodetector (PD) (6 dBm), and the signal is split between the PD (99\%) and an optical spectrum analyser (OSA).
The electrical signal from PD is transferred to a sampling oscilloscope (OSC).

\subsection{Light in- and out-coupling}

\begin{figure}[h!]
\centering
\includegraphics[width=0.7\linewidth]{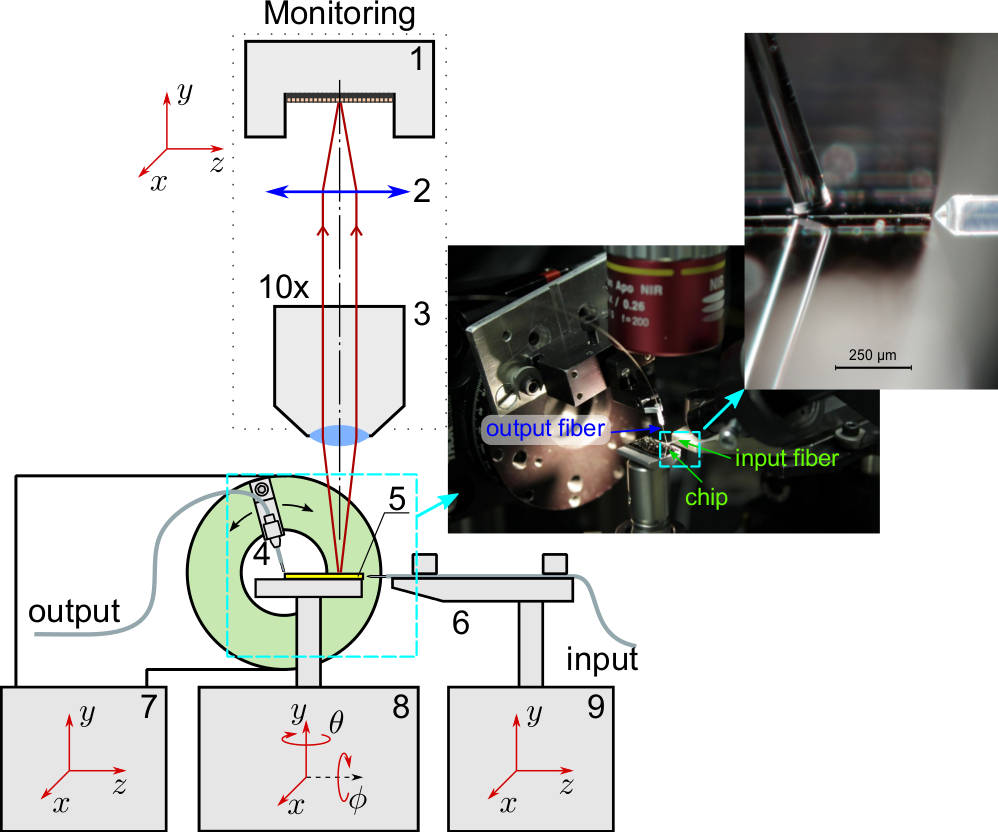}
   \caption{Setup for the light injection into a nanooptical waveguide.
    \emph{1}~--- infrared CCD camera;
    \emph{2}~--- tube lens;
    \emph{3}~--- 10$\times$ Mitutoyo infrared objective;
    \emph{4}~--- collecting cleaved fiber fixed in a ferrule on a rotating platform;
    \emph{5}~--- sample under test;
    \emph{6}~--- tapered optical fiber for incoupling clamped in a V-groove;
    \emph{7, 9}~--- translation stages;
    \emph{8}~--- translation/rotation stage. 
    \emph{Insets}: photo of optomechanical arrangement for in/out coupling; side view of the waveguide chip and input (lensed) and output (cleaved) fibers, obtained with VIS observation camera}
\label{fig:coupling_setup}
\end{figure}

The DUT includes, besides the test waveguide, a corresponding light coupling system (Fig.~\ref{fig:coupling_setup}). An input signal is coupled into the facet of waveguide using a tapered and lensed optical fiber. The output signal is collected with a cleaved optical fiber placed in the vicinity of the grating, which terminates the waveguide.

 Piezo-driven stages \emph{7--9} are used for a precise alignment of optical fibers and the sample. The lensed fiber for in-coupling is held in a V-groove \emph{6} and clamped with rare-earth magnets. The collecting cleaved fiber is glued into a ceramic ferrule, which is fastened up in a holder on a rotating platform. Such configuration allows us to place the fiber facet near the scattering grating and vary the collecting angle, which should be about 15$^\circ$.

The sample is illuminated by a pair of incandescent fiber optic illuminators and is observed with two digital cameras. An infrared camera from company Xenics \emph{1} is attached on top of a microscope, which consists of a 10$\times$ Mitutoyo IR objective \emph{3} (working distance of 32 mm) and a 200\,mm achromat tube lens \emph{2}.

Another monitoring tool is the Canon EOS 60D single-lens reflex photocamera mounted with the Canon EF MP-E 65\,mm macrolens that provides 5$\times$ magnification. The photocamera is used to observe the sample from a side (see Fig.~\ref{fig:coupling_setup}, inset). It is placed on a motorized linear translation rail (StackShot from Cognisys Inc.), which allows precise focusing. The purpose of the photocamera is to estimate the elevation of optical fibers above the surface of the sample.

\begin{figure}[h!]
\centering
\includegraphics[width=1\linewidth]{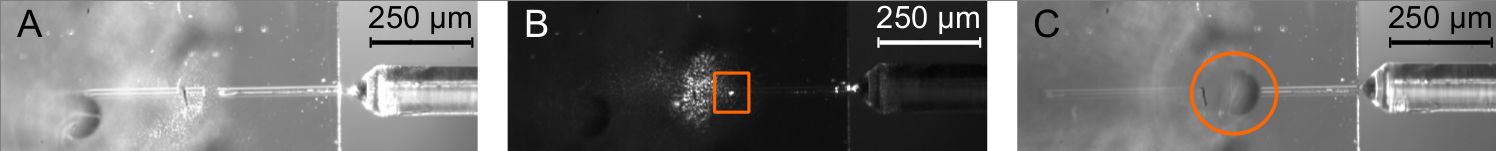}
 \caption{Images from the infrared camera during the
             alignment of fibers with respect to the waveguide.
             \emph{A}~--- adjustment of the illumination sources and rough
                         alignment of the tapered fiber (on the right);
             \emph{B}~--- fine adjustment of the tapered fiber until the strong
                         scattering is observed on the grating
                         (in the orange rectangle);
             \emph{C}~--- placement of the cleaved collecting fiber (highlighted with the orange circle) above
                         the grating}
\label{fig:coupling_alignment_IR}
\end{figure}

The fiber alignment procedure starts from the positioning of two light sources in a such way, that both the waveguides and the gratings are clearly visible (see Fig.~\ref{fig:coupling_alignment_IR}A). Light, which is irradiated in the transversal direction, gets reflected from the edges of the waveguides, whereas light going longitudinally scatters from the gratings.

After the waveguides are clearly seen, the tapered fiber is positioned in the vicinity of the waveguide facet (see Fig.~\ref{fig:coupling_alignment_IR}A). Next, the intensity of the illumination is reduced and a weak IR signal is sent through the tapered fiber. The latter one is aligned until one can see a strong emission from the grating (illustrated in Fig.~\ref{fig:coupling_alignment_IR}B).

Then the cleaved fiber is brought into the vicinity of the grating (see Fig.~\ref{fig:coupling_alignment_IR}C) and is moved down under the monitoring from the VIS camera. The intensity of the IR signal is increased up to 20 dBm, and coupling is optimized to achieve a maximal output signal.

\subsection{Evaluation of bit error rate}

\begin{figure}[h!]
     \centering
     \includegraphics[width=\textwidth]{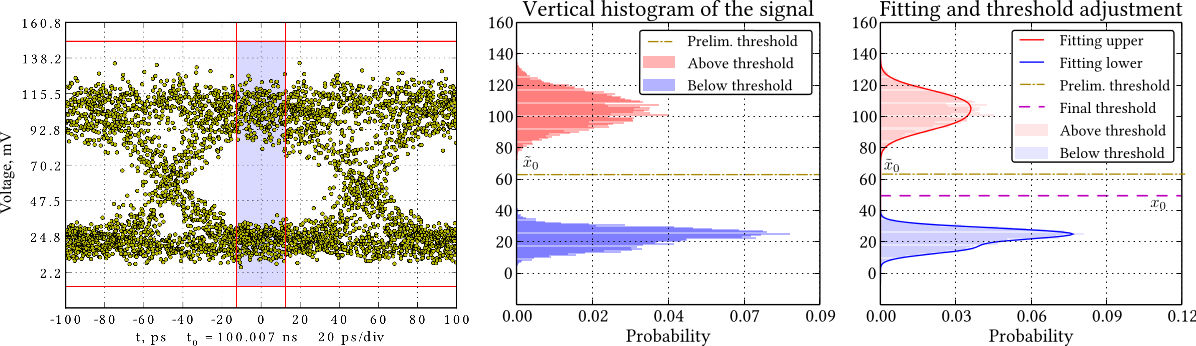}
     \caption{Estimation of PDF for received logical \texttt{ones} and \texttt{zeros}. \emph{Left:} the signal waveform is centred, and a histogram acquisition window is specified (blue rectangle).  \emph{Middle:} the OSC collects vertical histogram in the specified region. The middle plot shows the normalized histogram corresponding to the presented waveform.  \emph{Right:} fitting of each semi-histograms with a sum of two Gaussians (right plot). The intersection point of upper and lower fitting functions represents the decision threshold $x_0$}
     \label{fig:hist_processing_diagram}
    \end{figure}

The bit error rate (BER) is evaluated from eye diagrams, captured by a sampling oscilloscope.
The device acquires a considerable amount of waveforms and uses them to calculate a histogram of a signal voltage. This histogram represents the statistics of data hits in the specified window, anchored to the center of the eye (see Fig.~\ref{fig:hist_processing_diagram}). For the sufficiently large number of hits in the histogram window ($\sim 10^5$ hits), the normalized histogram approximates the probability distribution function (PDF) of the received electrical signal~-- logic ones and zeros.
    
    The histogram must be normalized before one can extract any statistics from it. If the histogram is represented as $N(x)$, where $N$ is the number of hits in a small interval $\Delta x$ and $x$ is the instantaneous voltage, then the normalized histogram $n(x)$ is found as:
    \begin{equation}
     n(x) = \dfrac{N(x)}{\Delta x \sum{N(x)}}.
     \label{eq:normalization}
    \end{equation}
    This results in a unit area under the histogram given by $\Delta x \sum n(x) = 1$ formula.

    The averaged voltage value is used as a preliminary decision threshold $\tilde{x}_0$:
    \begin{equation}
     \tilde{x}_0 = \Delta x \sum x \cdot n(x).
    \end{equation}
    It is used to split the histogram into upper $n_\uparrow(x)$ (corresponding to the binary \texttt{one}) and lower $n_\downarrow(x)$ (binary \texttt{zero}) parts.
    
    The PDFs of \texttt{one} and \texttt{zero} are close to Gaussian ones, but under the certain conditions a pair of close peaks is observed in each semi-histogram. Therefore, we perform fitting of the semi-histogram with a sum of two Gaussians (see Eq.~\ref{eq:fit_upper}).
The selection of an appropriate starting estimation plays a crucial role for a good fitting. For this reason we specify fitting functions $f_\uparrow(x)$ and $f_\downarrow(x)$ based on the statistics extracted from semi-histograms. The fitting function for logical \texttt{one} is:
    \begin{equation}
     f_\uparrow(x) = \frac{1}{2\sigma_\uparrow\sqrt{\pi}}\left(
                     \exp\left[-\frac{\left(x-\mu_\uparrow + \sigma_\uparrow\right)^2}
                                     {\sigma_\uparrow^2}\right] +
                     \exp\left[-\frac{\left(x-\mu_\uparrow - \sigma_\uparrow\right)^2}
                                     {\sigma_\uparrow^2}\right]\right),
		\label{eq:fit_upper}
    \end{equation}
where $\mu_\uparrow$ is the mean value and $\sigma_\uparrow$ is the standard deviation of data in the upper semi-histogram, which is supposed to be close to the normal distribution function.
    \begin{align}
     \mu_\uparrow &= \Delta x \sum{x \cdot n_\uparrow(x)}, \\
     \sigma_\uparrow &= \sqrt{\frac{\sum n_\uparrow(x) \cdot ( x - \mu_\uparrow)^2}{\sum n_\uparrow(x)}}.
    \end{align}
    The analogous relations are valid for the $f_\downarrow(x)$ function.

    The fitting functions $f_\uparrow(x)$ and $f_\downarrow(x)$ intersect with each other at the decision threshold value $x_0$. Therefore, the $x_0$ is found as a numerical solution of equation:
    \begin{equation}
     f_\uparrow(x_0) - f_\downarrow(x_0) = 0.
    \end{equation}

BER value is defined by a numerical integration of the overlapped regions according to formula:
    \begin{equation}
     \textnormal{BER} = \int\limits_{-\infty}^{x_0} f_\uparrow(x) \, dx +
                    \int\limits_{x_0}^{+\infty} f_\downarrow(x) \, dx.
    \end{equation}

\subsection{Estimation of BER penalties}
Each experiment is performed as following:
\begin{enumerate}
\item LR-DLSPPW is placed into the coupling setup (together they form DUT, item \emph{9} in Fig.~\ref{fig:experimental_setup}), and the BER at different values of received optical power (ROP) is determined. 
\item The DUT is replaced by the optical attenuator (VOA$_1$, item  \emph{10} in Fig.~\ref{fig:experimental_setup}), which provides the same value of insertion losses as DUT does (typically of about --42\,dB), and a set of reference back-to-back  measurements (B2B) is carried-out. During the B2B test the BER vs. ROP values are determined in the same way as described above.

\item Q-factor of the signal is retrieved from BER, according to the inverse function $Q=f^{-1}(\textnormal{BER})$, where $f^{-1}$ is calculated from \cite{Marcuse1990}:
\begin{equation}
\textnormal{BER}=f(Q) =\frac{1}{2}\textnormal{erfc}\left(\frac{Q}{\sqrt{2}}\right).
\end{equation}
\item The fitting of Q-factor vs. ROP is performed using the equation:
\begin{equation}
Q_\textnormal{fit}(x) = \frac{c_1 x}{\sqrt{c_2x^n+c_3}+c_4}+c_5, \quad n=2.
\label{eq:Q_factor_fit}
\end{equation}

Derivation of Eq.~\ref{eq:Q_factor_fit}, with $n=1$, is provided elsewhere \cite{Marcuse1990}. The power $n=2$ has been introduced in the current fitting function to take into account the saturation of Q-factor at high ROP values.

\item The BER$(Q_\textnormal{fit})$ dependency is evaluated using Eq.~\ref{eq:Q_factor_fit} both for DUT and B2B tests.

\item BER penalties, which characterize the data signal degradation in the presence of DUT,  are defined as:
\begin{equation}
\textnormal{PEN}=\left|\log\dfrac{\textnormal{BER}(Q_\textnormal{fit}^\textnormal{DUT})}{\textnormal{BER}(Q_\textnormal{fit}^\textnormal{B2B})}\right|.
\end{equation}

\end{enumerate}

\section{Results}
The performed experiments on data transmission can be divided into subgroups, which differ by:
\begin{enumerate}
\item Number of transmitted channels;
\item Channel wavelengths;
\item Spacing between channels.
\end{enumerate}

The number of channels is limited by the total gain of post-amplifiers (items \emph{11}  and \emph{13} in Fig.~\ref{fig:experimental_setup}). As the total power of the injected signal should be kept constant, adding $N$ extra channels to the system involves $10 \log N$ dB reduction of the power level of every channel. Thus, the gain of the post-amplifies should be increased by the same amount to maintain the channel power constant. From the other side, an input power to the waveguide is increased with the number of channels, which finally can lead to the thermal damage to the waveguide.

\begin{figure}[h!]
\centering
\includegraphics[width=0.8\linewidth]{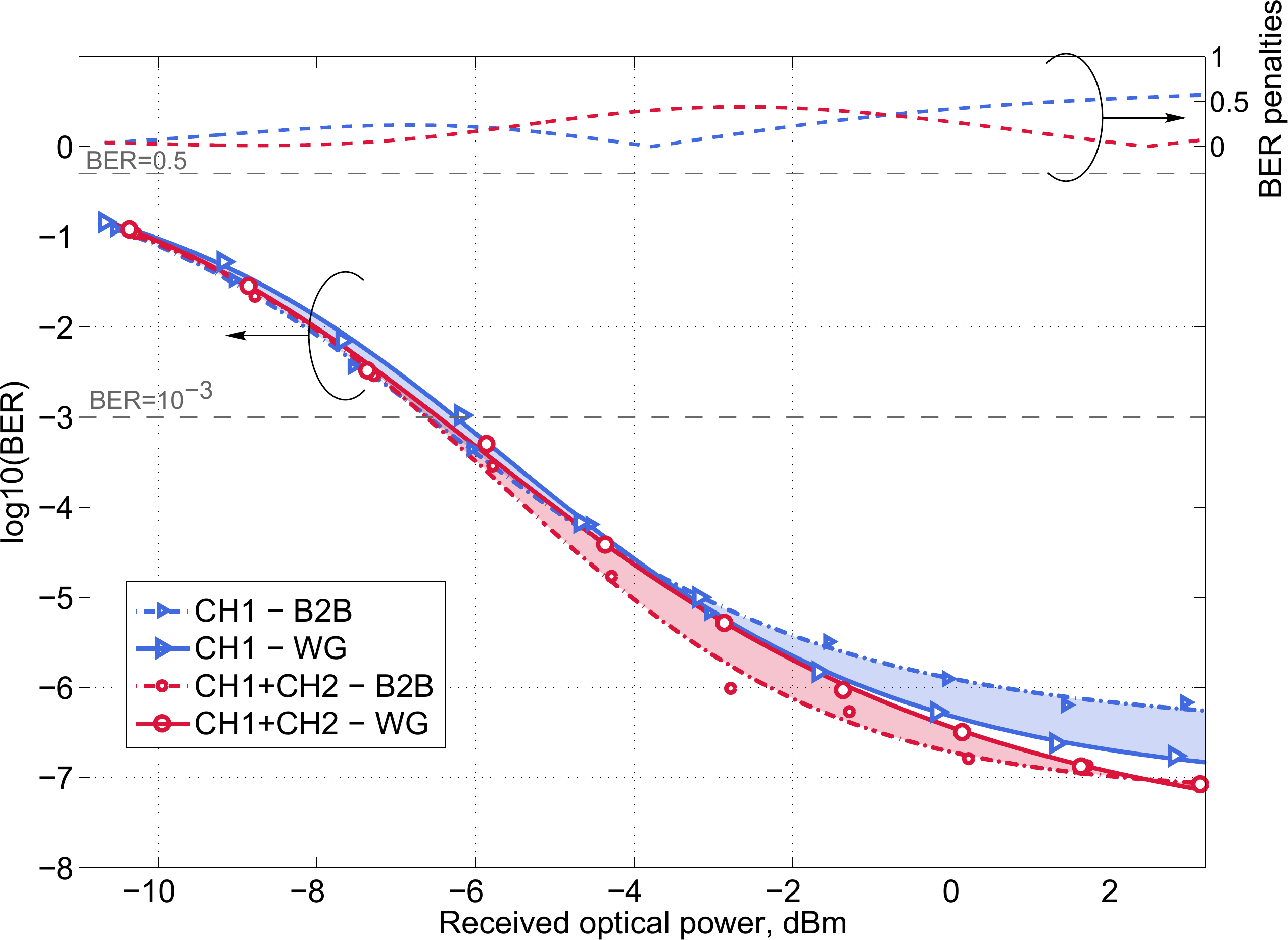}
\caption{BER and penalties for single and double channel transmission at 1549 nm, channel spacing 0.4 nm: \emph{B2B} -- back-to-back measurements, \emph{WG} -- measurements on LR-DLSPPW}
\label{fig:pen_single_double_ch_1549_nm}
\end{figure}

As it was described in the previous section, one or two channel signals have been transmitted through the LR-DLSPPW.
Results of experiments with single and double channel (0.4 nm spacing) data transmission at 1549 nm are presented in Fig.~\ref{fig:pen_single_double_ch_1549_nm}. It should be noted that the BER penalties remain below 0.5 dB in the ROP range of interest from $-6.6$ to $3$ dBm (lower detected powers results in the BER that exceeds threshold $10^{-3}$ required for the proper operation of forward error correcting codes (FEC)  \cite{Lize2007, Weber2010, Hood2012}). Moreover, we managed to achieve BER as low as $10^{-7}$.

\begin{figure}[h!]
\centering
\includegraphics[width=0.8\linewidth]{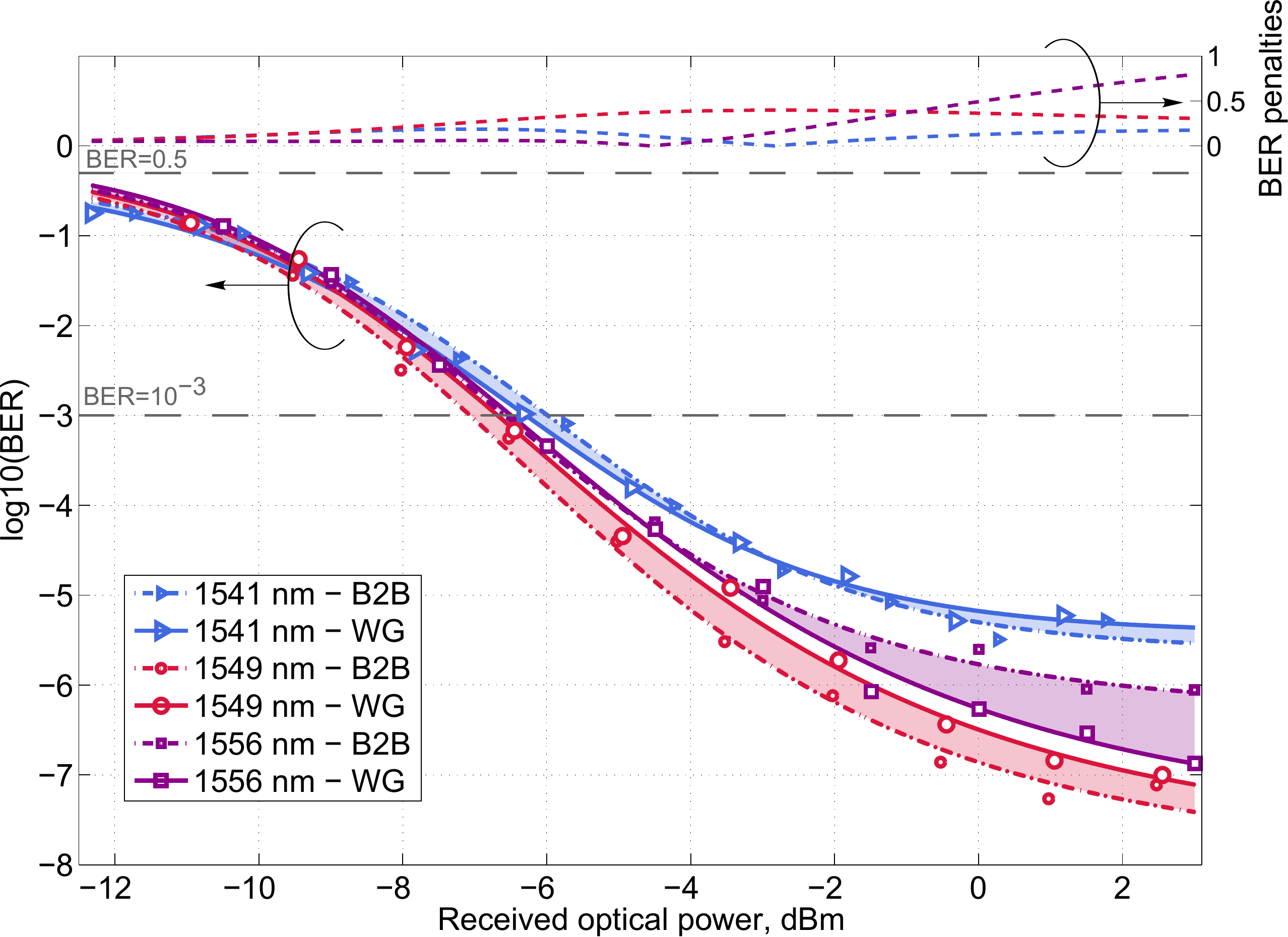}
\caption{BER and penalties for double channel transmission in different wavelength regions, channel spacing 0.4 nm: \emph{B2B} -- back-to-back measurements, \emph{WG} -- measurements on LR-DLSPPW}
\label{fig:double_ch_diff_wav}
\end{figure}

The BER dependence on the transmitted wavelength has been studied in three different wavelength regions (Fig.~\ref{fig:double_ch_diff_wav}). As in the previous case, BER penalties do not exceed 0.6 dB at ROP from $-7$ to $3$ dBm. The full wavelength region under study was limited by the gain profile of exploited EDFAs. In some cases back-to-back measurements turned out to be a bit worse than the respective ones with the waveguide. We intend to attribute these discrepancies to the adopted in this paper method of the BER evaluation.

\begin{figure}[h!]
\centering
\includegraphics[width=0.8\linewidth]{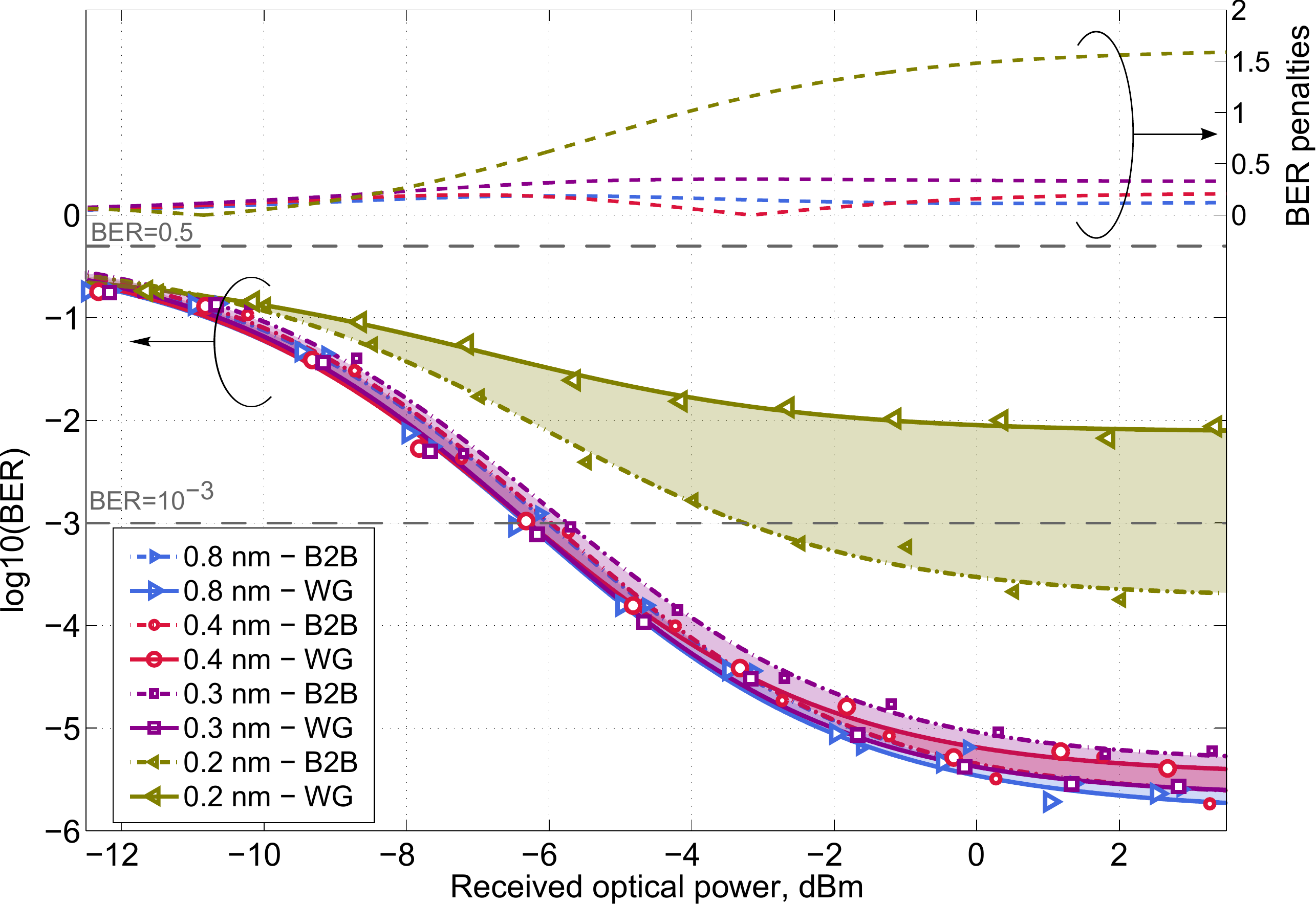}
\caption{BER and penalties for double channel transmission at 1541 nm, different inter-channel spacing: \emph{B2B} -- back-to-back measurements, \emph{WG} -- measurements on LR-DLSPPW.}
\label{fig:double_ch_diff_spacing}
\end{figure}

Finally, we evaluated BER penalties originating from the signal crosstalk by changing the spectral distance between channels (Fig.~\ref{fig:double_ch_diff_spacing}). BER penalties below 0.5 dB are obtained for the wavelength spacing as low as 0.3 nm ($\sim37\,\unit{GHz}$). However, penalties of nearly 1.5 dB are estimated for the channel spacing of 0.2 nm ($\sim25\,\unit{GHz}$), indicating the presence of channel crosstalk in the LR-DLSPPW. The appeared cross talk at the channel separtion of 2\,nm is caused by a band pass filter that was placed before the receiver. The purpose of the filter is to select a single channel; however, the filter bandwidth is about 0.2\,nm, which is comparable with the channel separation, so the receiver additionally detects signal from the wrong channel.

\newpage
\section{Summary}

In conclusion, we have demonstrated the data transmission of 10 Gbit/s on-off keying modulated 1550~nm light wave through a LR-DLSPPW structure with negligible signal degradation. The BER penalties do not exceed 0.6~dB over the 15~nm wavelength range and ROP between $-7$ and $3$~dBm. The BER penalties were determined by the comparison of LR-DLSPPW structure with an optical attenuator that provides the same insertion loss. We should note, however, that the main contributor to the observed insertion loss is the light scattering at the interfaces between waveguide facets and the optical fibers of the testbench.

The achieved results show the applicability of LR-DLSPP waveguiding structures for the data transmission in integrated photonic interconnects. It demonstrates long propagation distances together with the subwavelength mode confinement; recent studies \cite{Volkov2011} show that the waveguide could be patterned along curved lines or even be split in two. Moreover, authors of \cite{grandidier2009} present techniques for loss compensation in LR-DLSPPW modes by means of stimulated emission. Such unique combination of features holds promise for implementation of compact and powerful integrated plasmonic circuits for fast photonic data processing devices.

\newpage
 \printbibliography
\end{document}